\documentstyle[amsfonts,aps,preprint]{revtex}	
\begin{document}
%
%
\preprint{EC-HEP-950726}
\date{July 26, 1995}
\title{Di-photon + Jet Event Structure}
\author{B.~Bailey}
\address{
Department of Physics,
Eckerd College,
Saint Petersburg, Florida 33733
}
\maketitle
\begin{abstract}
New distributions
are presented which allow di-photon + jet events to be
clearly separated into three classes of events based on the $p_{T}$
of the final state particles and their separation
$R$ = $\sqrt{(\Delta y)^{2} + (\Delta \phi)^{2}}$.
The analysis used can easily be extended to the case of di-jet + photon.
\end{abstract}
\pacs{PACS numbers: 12.38.Bx, 14.80.Er}
%
%
\widetext

A recent calculation of di-jet + photon suggested a method of
separating these events into three classes based on whether the
photon had the highest, middle or lowest energy fraction~\cite{JK}.
I refer the interested reader to Ref.~\cite{JK} and references
contained therein. The analysis here uses different observables
and distributions. The cuts necessary are simple and the observables
may be constructed from the four-vectors of the final state particles
in the lab (hadron-hadron center-of-momentum) frame.

The di-photon + jet cross section consists of, at the Born
level, $q \bar{q} \rightarrow \gamma \gamma g$,
$q g \rightarrow \gamma \gamma q$ and
$\bar{q} g \rightarrow \gamma \gamma \bar{q}$.
In this brief note it will be shown that by ordering events by the
$p_{T}$ of the jet, and looking at experimentally
motivated distributions, that di-photon + jet events can be
clearly separated into three classes. These three classes are: (1)
$p_{T jet}>$ $p_{T \gamma 1}>$ $p_{T \gamma 2}$,
(2) $p_{T \gamma 1}>$ $p_{T jet}>$ $p_{T \gamma 2}$,
and (3) $p_{T \gamma 1}>$ $p_{T \gamma 2}>$ $p_{T jet}$. The distributions
in question are motivated by the cone algorithm,
$R$ = $\sqrt{(\Delta y)^{2} + (\Delta \phi)^{2}}$, used to define jets
and isolated photons. Where $\Delta y$ is the magnitude of the
difference in rapidity of two
final state particles and $\Delta \phi$ is the azimuthal opening angle
between the two final state particles.
The distributions which will be presented in this paper
are: ($1/\sigma$) $d \sigma/d(\Delta y_{\gamma \gamma})$,
($1/\sigma$) $d \sigma/d(\Delta \phi_{\gamma \gamma})$
and ($1/\sigma$) $d \sigma/dR_{\gamma \gamma}$.
The distributions are normalized by the cross section to reduce the
sensitivity to $Q^{2}$ and differences in parton distributions.
The following inputs are used for this calculation:
$\sqrt{s}$ = 1800 GeV,
CTEQ3L parton distributions~\cite{CTEQ3}, the one-loop expression for
$\alpha_{s}(Q^{2})$, $Q^{2}=p^{2}_{T \gamma}$, and $m_{top}=$ 180 GeV.
Additionally, unless otherwise stated, the following cuts
are used: $p_{T \gamma}>$~10 GeV, $p_{T jet}>$~10 GeV,
$|y_{\gamma}|<$ 2.5, $|y_{jet}|<$ 3.0, and $R >$ 0.4.
%
%

Fig.~1 shows the results for the $\Delta y_{\gamma \gamma}$
distribution. The dashed curve is class (1), dot-dash class (2),
dotted class (3), and the solid curve is the sum of the three. It can be
seen that the distribution is a rapidly falling function of
$\Delta y_{\gamma \gamma}$ and that the three classes are distinct.
It should be noted that the separation can be enhanced by requiring a
stiffer $p_{T}$ cut on the photons thus enhancing class (3) events
at the cost of reducing the total number of events.

%
%
Fig.~2 shows the results for the $\Delta \phi_{\gamma \gamma}$
distribution. This distribution has three distinct features: a
``shoulder'', a ``double-hump'', and a ``bump''. Each is associated with
class (1), class (2) and class (3) type events accordingly.
The locations of each peak can be understood easily from the kinematics
of each class.
In class (1) events both photons have lower $p_{T}$
than the jet and are thus recoiling against the jet. Therefore class (1)
events dominate at low $\Delta \phi_{\gamma \gamma}$.
The probability to have a class (1) event increases with
$\Delta \phi_{\gamma \gamma}$ until $\Delta \phi_{\gamma \gamma}$
reaches $\frac{\pi}{2}$ and class (2) events ``turn on''.
The ``double-hump'' or twin-peak structure of class (2) events can also be
understood. In class (2) events the jet $p_{T}$ is intermediate
in the $p_{T}$ ordering. The probability of class (2) events also
increases with $\Delta \phi_{\gamma \gamma}$ to peak at $\frac{2\pi}{3}$.
These events are the so-called ``Benz'' events where each of the final
state particles has nearly equal momentum. This is also the location of
the ``turn on'' of class (3) events.
Both class (2) and class (3) events
are strongly peaked near $\pi$. For class (2) events this occurs when one
of the photons is soft and in class (3) events when the jet is soft.
It is important to remember that although kinematics determines the location
of the various peaks, the matrix elements determine the relative heights
and shapes of the distributions.
Also, as in the $\Delta y_{\gamma \gamma}$ distribution,
the separation of the classes can be enhanced by requiring a
stiffer $p_{T}$ cut on the photons thus enhancing class (3) events
at the cost of reducing the total number of events.

%
%
Fig.~3 shows the results for the $R_{\gamma \gamma}$
distribution. The three classes are once again clearly separated and
retain much of the structure shown in the $\Delta \phi_{\gamma \gamma}$
plots. The ``shoulder'' from class (1) events is clearly visible. The
``double-hump'' of class (2) events has been smeared out, but the peaking
of class (2) and class (3) events at $R$ near three is a remnant of
$\Delta \phi_{\gamma \gamma}$ peaking near $\pi$.
Also, as in the previous distributions,
the separation of the classes can be enhanced by requiring a
stiffer $p_{T}$ cut on the photons thus enhancing class (3) events
at the cost of reducing the total number of events.

{}From a experimental point of view it should be noted that the requirements on
the jet can be relaxed. In fact the jet need not be directly observed, just
inferred from a large $p_{T}$ imbalance between the photons. This removes the
uncertainties associated with jet definitions, jet momentum and jet
triggering efficiencies. As mentioned in the text, the various
classes of events can be enhanced by adjusting the $p_{T}$ cut on the photons
or the jet. It should also be noted that the photons can be restricted to
more central rapidities at the cost of a lower cross section.
Finally it should be noted the analysis presented here can
be carried over to the di-jet + photon process
(jets and photon must be directly observed).

%
\acknowledgements
This research was supported in part by Eckerd College and is a continuation
of research that was supported by the U.~S. Department of Energy under
contract number DE-FG05-87-ER40319. The author would like to
thank the Florida State University Department of Physics and Argonne
National Laboratory for the use of computer facilities. Additionally,
the author would like to thank G. Blazey and J. Womersley for stimulating
discussions.
%
%

%
%
%
\figure{Figure 1. $d \sigma/d(\Delta y_{\gamma \gamma})$
vs. $\Delta y$ distribution using the cuts described
in the text. The dashed curve is class (1), dot-dash curve is class (2),
dotted curve is class (3), the solid curve is the sum of the three.
\label{FIG 1} }
%
\figure{Figure 2. $d \sigma/d(\Delta \phi_{\gamma \gamma})$
vs. $\Delta \phi$ distribution
using the cuts described
in the text. The dashed curve is class (1), dot-dash curve is class (2),
dotted curve is class (3), the solid curve is the sum of the three.
\label{FIG 2} }
%
\figure{Figure 3. $d \sigma/dR_{\gamma \gamma}$ vs. $R$ distribution using
the cuts described in the text.
The dashed curve is class (1), dot-dash curve is class (2),
dotted curve is class (3), the solid curve is the sum of the three.
\label{FIG 3} }
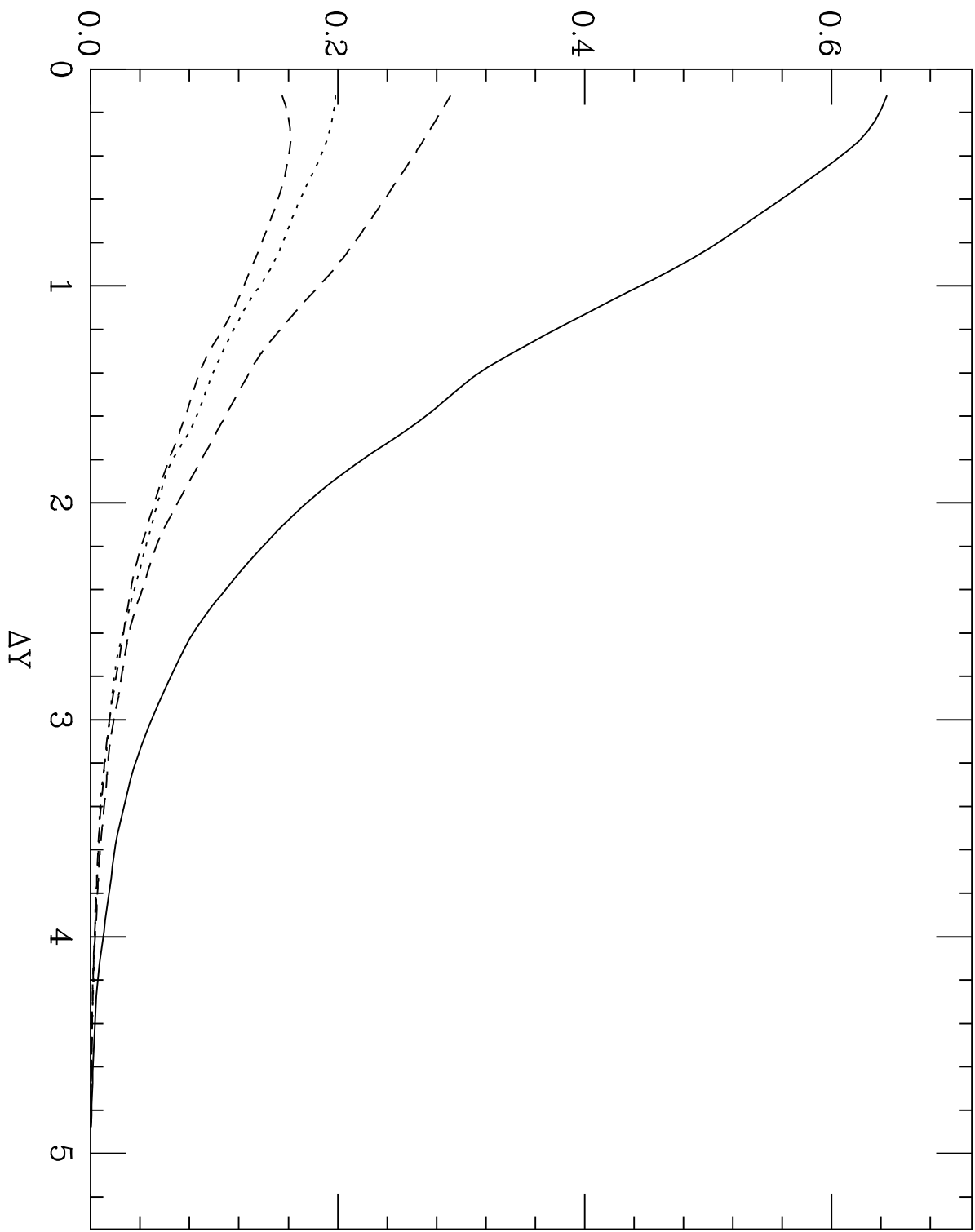
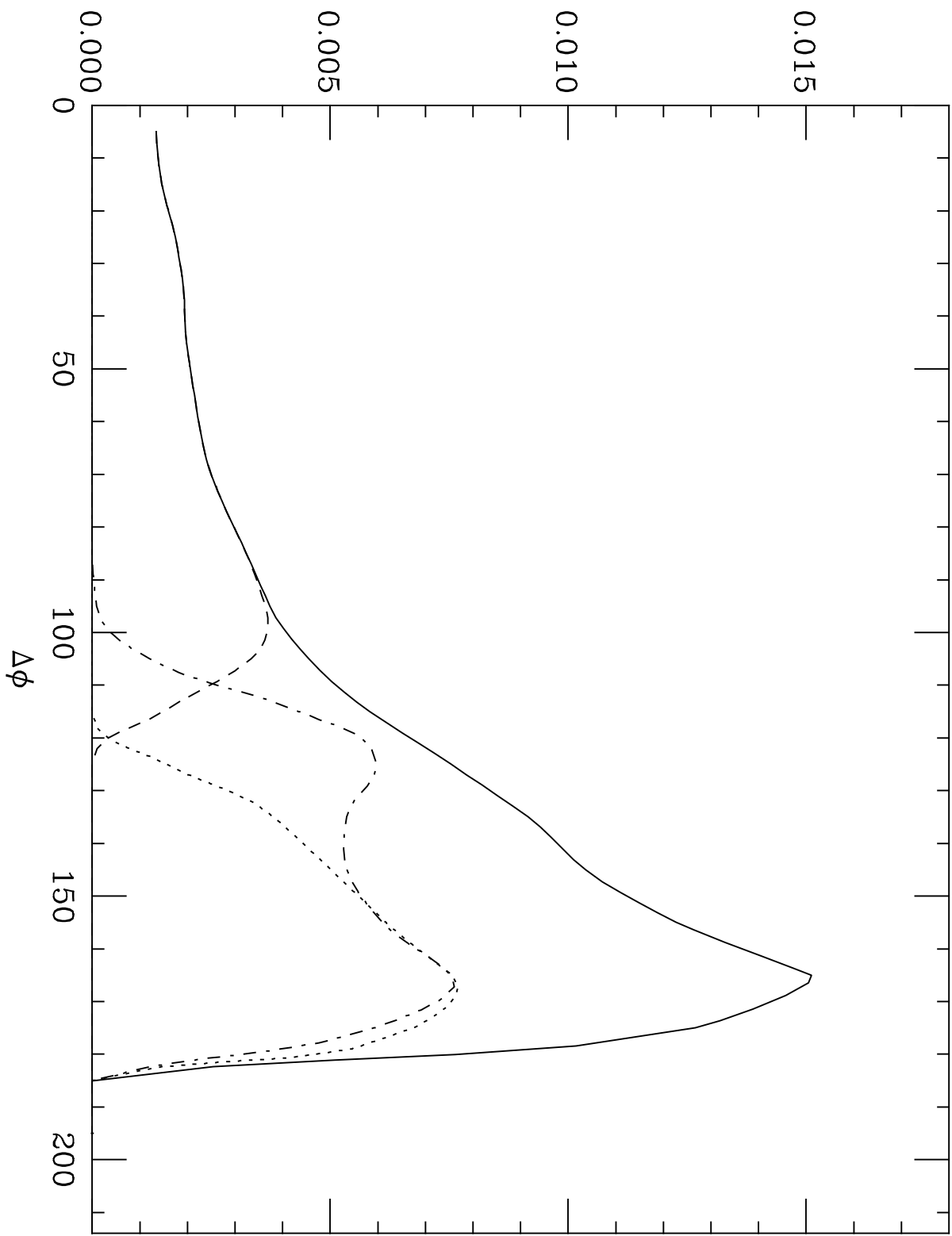
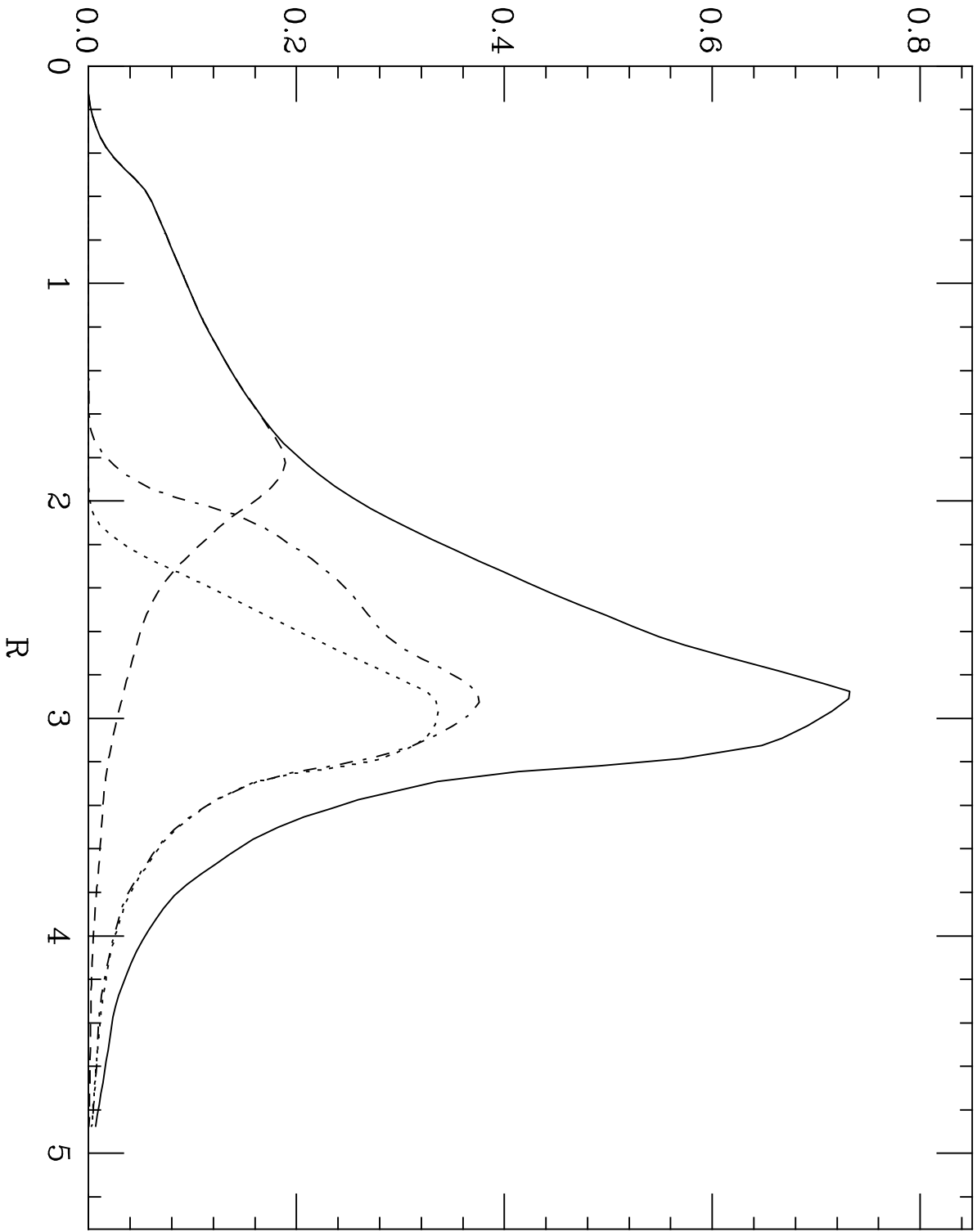
\end{document}